\documentclass{aa}

\usepackage{graphicx}
\usepackage{times}

\begin{document}

   \thesaurus{06         
              (13.07.1)}  
            \title{Observations of short-duration X-ray transients by WATCH
                   on Granat}



   \author{A. J. Castro-Tirado
          \inst{1,2}
   \and S. Brandt
          \inst{3}
   \and N. Lund
          \inst{4}
   \and R. Sunyaev
          \inst{5}
          }

   \offprints{A. J. Castro-Tirado (ajct@laeff.esa.es)}

   \institute{Laboratorio de Astrof\'{\i}sica Espacial y F\'{\i}sica 
              Fundamental (LAEFF-INTA), 
              P.O. Box 50727, E-28080 Madrid, Spain
   \and Instituto de Astrof\'{\i}sica de Andaluc\'{\i}a (IAA-CSIC),
              P.O. Box 03004, E-18080 Granada, Spain
   \and Los Alamos National Laboratory, Los Alamos, NM 87545, USA
   \and Danish Space Research Institute, Marianne Julies Vej 30, 
        DK-2100, Copenhagen O, Denmark
   \and Space Research Institute, Russian Academy of Sciences,
        Profsouznaya 84/32, 117810 Moscow, Russia
             }

   \date{Received date; accepted date}

  \titlerunning{Short-duration X-ray transients}

  \authorrunning{Castro-Tirado et al.}

   \maketitle

   \begin{abstract}
During 1990-92, the WATCH all-sky X-ray monitor on {\it Granat} discovered six 
short-duration X-ray transients. In this paper we discuss their possible 
relationship to peculiar stars. 
Only one of the fast (few hours) X-ray transients (GRS 1100-771) might be 
tentatively ascribed to a superflare arising from a young stellar object in 
the Chamaeleon I star-forming cloud.
At the distance of $\sim$ 150 pc, L$_{x}$ = 1.35 $\times$ 
10$^{34}$  \rm \ erg \ s$^{-1}$  (8-15 keV), or  
2.6 $\times$ 10$^{34}$  \rm \ erg \ s$^{-1}$  (0.1-2.4 keV)
assuming a thermal spectrum with kT $\sim$ 10 keV, a temperature  
higher than those previously seen in T Tauri stars (Tsuboi et al. 1998). 
The {\it peak} X--ray luminosity is at least 2 times higher than that 
derived for the protostar IRS 43 (Grosso et al. 1997) which would make 
-to our knowledge- the strongest flare ever seen in a YSO. 
However, the possibility of GRS 1100-771 being an isolated neutron star 
unrelated to the cloud cannot be excluded, given the relatively large error 
box provided by WATCH.
Regarding the longer duration
($\sim$ 1 day) X-ray transients, none of them seem to be related to known
objects. We suggest that the latter are likely to have originated from compact 
objects in low-mass or high-mass X-ray binaries, similarly to XTE J0421+560.

     \keywords{X-rays: bursts --- X-rays: stars --- Stars: variables: general 
               --- Stars: flare --- Stars: pre-main sequence 
               --- ISM: Chamaeleon I cloud}

   \end{abstract}

%

\section{Introduction}

Amongst the many kinds of sources in the variable X-ray sky, X-ray transients 
have been observed since the first experiments in the late 60's. According to 
their duration, it is possible to distinguish between Long-duration X-ray 
Transients (lasting from weeks to few months) and Short-duration X-ray 
Transients (lasting from hours to very few days).

Long-duration X-ray Transients are mainly related to Be-neutron star systems 
(Be X-ray Transients) and to Soft X-ray Transients (a subclass of low-mass 
X-ray binaries). The latter include the best black hole candidates found so 
far (Tanaka \& Shibazaki 1996).

But not too much is known with respect to most of Short-duration X-ray 
Transients. The main reason is the lack of counterparts at other wavelengths.
One subclass are the so-called Fast X-ray Transients, that have been observed 
with many detectors since the launching of the {\it Vela} satellites 
(Heise et al. 1975, Cooke 1976) until the advent of the {\it BeppoSAX} 
satellite (Heise et al. 1998). 
Durations range from seconds (Belian, Conner \& Evans 
1976) to less than few hours. Normally, they have been seen once, and never 
seen in quiescence, implying high peak-to-quiescent flux ratios 
(10$^{2}$-10$^{3}$) (Ambruster et al. 1983). 
Spectral characteristics vary substantially, from a hard spectrum in MX 2346-65
(kT $\sim$ 20 keV; Rappaport et al. 1976) to soft spectra with blackbody 
temperatures from kT = 0.87 to 2.3 keV (Swank et al. 1977). In two cases, 
precursors to the main event were observed by {\it SAS 3} (Hoffman et al. 
1978). The 
precursors rose and felt in brighteness in less than 0.4 s. In the Ariel V 
database, 27 sources were discovered (Pye \& McHardy 1983), and 10 more were 
detected in the {\it HEAO 1} A-1 all sky survey (Ambruster \& Wood, 1986), 
implying
a fast transient all-sky event rate of $1500\rm\ yr^{-1}$ for fluxes F$_{x}$ 
$\geq$  3 $\times$ 10$^{-10}$ erg cm$^{-2}$ s$^{-1}$ in the 2-10 keV 
energy band. In the {\it HEAO 1} A-2 survey, 52 events were found 
(Connors 1988), but 37 of them were related to four of the brightest 
X-ray sources in the LMC.

Due to the large difference of observational characteristics, it seems that 
these events are caused by more than one physical mechanism. In several cases, 
there have been tentative optical identifications on the basis of known 
sources in the transient error boxes. One suggestion has been that many of the 
fast X-ray transients are related with stellar flares originated in active 
coronal sources, like RS CVn binaries or dMe-dKe flare stars.
RS CVn systems are binaries formed by a cool giant/sub\-giant (like a K0 IV)
with an active corona and a less massive companion (a late G-dwarf) in a 
close synchronous orbit, with typical periods of 1-14 d. Peak luminosities are 
usually  L$_{x}$ $\sim$  10$^{32}$ erg s$^{-1}$. The RS CVn system LDS 131 
was identified with the X-ray transient detected by {\it HEAO 1} on 9 Feb 1977 
(Kaluzienski et al. 1978, Griffiths et al. 1979).  The highest peak 
luminosity was recorded by {\it Ariel V} for the flare observed from 
DM UMa in 1975 
(Pye \& Mc Hardy, 1983). The hardest flare yet observed was for the system 
HR 1099, on 17 Feb 1980, which was detected by {\it HEAO 2} at energies up to 
20 keV (Agrawal \& Vaidya 1988). The most energetic X-ray flare was observed 
by {\it GINGA} from UX Ari. Its decay time was quite long ($\sim$ 0.5 days).

\begin{table*}[t]
\begin{center}
\caption{WATCH Fast X-ray Transients.}
  \begin{tabular}{cccccccc}
\hline
Source      &    date    & U.T. (onset)   &   duration ({\it min})       & $\alpha$(2000.0)  &
 $\delta$(2000.0) &  b  &  F$_{(\rm 8-15 ~keV)}$(erg cm$^{-2}$ s$^{-1}$)\\
\noalign{\smallskip}
\hline
\noalign{\smallskip}
GRS 1100-771   &    15 Jan 1992 & 13:39 & 90 &   11h 01m.2 &  -77$^{\circ}$.4   & -16$^{\circ}$.0 & 5.0 $\times$ 10$^{-9}$ \\
GRS 2037-404   &    23 Sep 1992 & 04:35 & 110 &   20h 41m.2 &  -39$^{\circ}$.6  & -38$^{\circ}$.0 & 4.7 $\times$ 10$^{-9}$ \\
GRS 2220-150   &    19 Sep 1990 & 03:32 & 240 &   22h 23m.1 &  -14$^{\circ}$.6  & -54$^{\circ}$.0 & 1.3$\times$ 10$^{-8}$ \\
\hline
\end{tabular}
\end{center}
\end{table*}

X-ray flares can be also observed from M or K dwarfs with Balmer lines in
emission (these are the active cool dwarf stars dMe-dKe). In the {\it Ariel 
V} sky 
survey, Rao \& Vahia (1984) suggested seven dMe stars as responsibles of X-ray 
flares that re\-ached peak luminosities L$_{x}$ $\leq$ 10$^{32}$ erg s$^{-1}$  
(2-18 keV). AT Mic is the dMe star with the largest number of recorded events 
(four), with a big flare in 1977 (Kahn 1979) reaching a peak luminosity  
L$_{x}$ = 1.6 $\times$ 10$^{31}$ erg s$^{-1}$  (2-18 keV), which is $\sim$
100 times 
larger that the strongest solar flares. One of the most energetic flares was 
the flare observed by {\it EXOSAT} in YY Gem on 15 Nov 1984. Pallavicini et 
al. (1990) estimated a total energy flare E$_{x}$ =  10$^{34}$ erg 
(0.005-2 keV), and a decay time t$_{d}$ =  65 min (one of the longest decay 
times ever mesured for such events). One of the longest flare ever reported 
was observed for more than 2 hours for the the X-ray source EXO 040830-7134.7 
(van der Woerd et al. 1989).

X-ray flares have been observed from Algol-type binaries (Schnopper et
al. 1976, Favata 1998), W UMa systems and young stellar objects (YSOs). Most
of the YSOs are deeply embedded young stars and T Tauri stars.
T Tauri stars are pre-main sequence stars (ages 10$^{5}$-10$^{7}$ yr) that 
may exhibit X-ray flaring activity via thermal bremsstrahlung from a hot 
plasma (see Linsky 1991). 
Montmerle et al. (1983) reported a "superflare" in the T Tauri star ROX-20, 
in the $\rho$ Oph Cloud Complex, which reached a peak luminosity  
L$_{x}$ = 1.1 $\times$ 10$^{32}$ \rm \ erg \ s$^{-1}$ (0.3-2.5 keV), 
and an integrated flare X-ray energy E = 10$^{34}$ \rm \ erg.
An  {\it ASCA} observation of a flare in V773 Tau implied  
L$_{x}$ $\sim$ 10$^{33}$ \rm \ erg \ s$^{-1}$ (0.7-10 keV),
and a total energy release of $\sim$ 10$^{37}$ \rm \ erg, which is
among the highest X-ray luminosities observed for T Tau stars (Tsuboi et al. 
1998). 
Recently, X-ray activity in protostars has been detected.
These are objects closely related to T Tau stars, with ages 10$^{4}$-10$^{6}$ 
yr, which are known to be strong X-ray sources when they enter the T Tauri 
phase.  See Montmerle et al. (1986) and Montmerle \& Casanova (1996) for 
comprehensive reviews. 
Koyama et al. (1996) reported a extremely high temperature 
(kT $\sim$ 7 keV) {\it quiescent} X-ray emission from a cluster of protostars 
detected by {\it ASCA} in the R CrA molecular cloud. Preibisch (1998) also 
reported a {\it ROSAT} observation of the source EC 95 within the 
Serpens star forming region, for which a {\it quiescent} (dereddened) 
soft X-ray luminosity
of $\sim$ (6-18) $\times$ 10$^{32}$ erg/s is derived, which is at least 60 
times larger than that observed for other 7 YSOs. Two of these YSOs have
displayed flaring activity. In particular, IRS 43 in the $\rho$ Oph cloud
showed an extremely energetic superflare with a {\it peak} luminosity of
L$_{x}$ = 6 $\times$ 10$^{33}$ \rm \ erg \ s$^{-1}$ (0.1-2.4 keV) (Grosso et
al. 1997). 

Theoreticians have suggested other sources like dwarf 
novae (Stern, Agrawal \& Riegler 1981) or bizarre type I X-ray bursts (Lewin 
\& Joss 1981) as origin of some fast X-ray transients.

\section{Observations}

The WATCH instrument is the all-sky X-ray monitor onboard the {\it Granat} 
satellite, 
launched on 1 Dec 1989. It is based on the rotation modulation collimator 
principle (Lund 1985). The two energy ranges are approximately 8-15 and 
15-100 keV. Usually, the uncertainty for the location of a new and short-duration
source is 1$^{\circ}$  error radius (3-$\sigma$). More details can be found in 
Castro-Tirado (1994) and Brandt (1994).

\subsection{Fast X-ray Transients discovered by WATCH}

During 1990-1992, WATCH discovered three bright fast X-ray transients. 
A small quantity when we compare with the other instruments mentioned above. 
This is mainly due to the higher low energy cut-off for WATCH ($\sim$ 8 keV), 
implying that only the harder events are detected. The events were immediately 
noticeable as an increase in the low energy count rate (Fig. 1) lasting up to
a few hours. No positive detections were made for the higher energy band. 
Their observational characteristics are summarized on Table 1.

\begin{figure}[!th]
    \resizebox{\hsize}{!}{\includegraphics{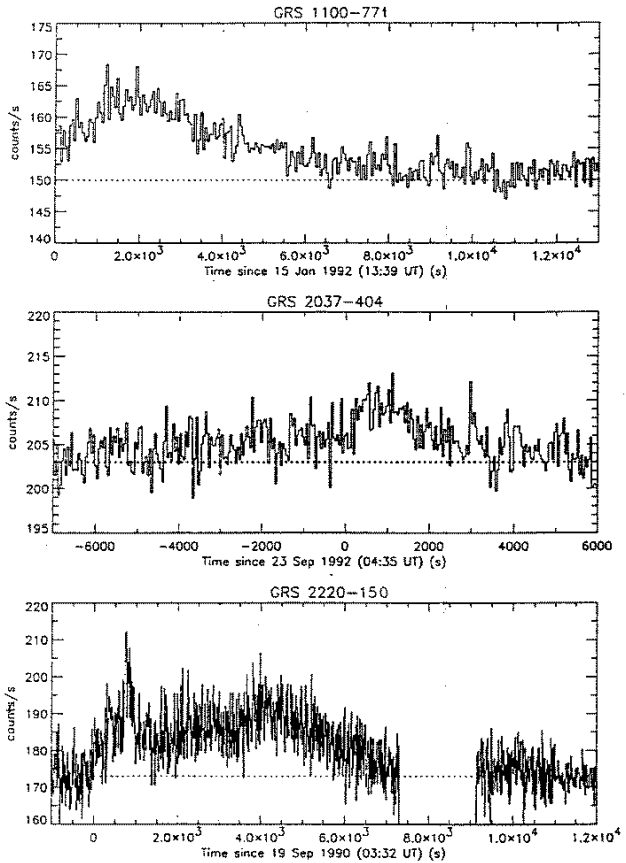}}
    \caption{The X-ray light-curves in the 8-15 keV of the three fast 
             X-ray transients discovered by WATCH on Granat.
             }
\end{figure}

\subsection{Longer Duration X-ray Transients discovered by WATCH}

Three objects were also discovered by WATCH during 1990-92, although it is 
likely that some others are present in the whole WATCH data base (1990-96). 
The sources reported here 
were observed to peak at X-ray fluxes up to 3 $\times$ 10$^{-9}$ 
erg cm$^{-2}$ s$^{-1}$ in the 8-15 keV band, and were 
discovered by analysis of the corresponding modulation 
pattern data. Their observational characteristics are summarized on Table 2.  
A fourth event, {\it GRS 1133+54}, lasting for $\sim$ 1 day on 19-20 Nov 1992 
was reported by Lapshov et al. (1992). 
        
\begin{table*}[ht]
\begin{center}
\caption{WATCH long-lived X-ray Transients.}
  \begin{tabular}{lccccccc}
\hline
Source      &    date    & U.T. (onset)   &   duration ({\it days})       & $\alpha$(2000.0)  &
 $\delta$(2000.0) & b & F$_{(\rm 8-15 ~keV)}$(erg cm$^{-2}$ s$^{-1}$) \\
\noalign{\smallskip}
\hline
\noalign{\smallskip}
GRS 0817-524  &   10 Oct 1990 & 14:00 & $\sim$ 1 &  08h 19m.4 &  -52$^{\circ}$.2 & -8$^{\circ}$.9 & 3.1 $\times$ 10$^{-9}$ \\
GRS 1133+540$^{a}$ &   19 Nov 1992 & 12:00 & $\sim$ 1 &  11h 35m.7 &  +53$^{\circ}$.7 & +60$^{\circ}$.4 & 2.2 $\times$ 10$^{-9}$ \\
GRS 1148-665  &   10 Oct 1990 & 12:00 & $\sim$ 1 &  11h 50m.4 &  -66$^{\circ}$.8 & -4$^{\circ}$.8 & 3.1 $\times$ 10$^{-9}$ \\
GRS 1624-375  &   24 Sep 1992 & 03:00 & $\sim$ 1 &  16h 27m.7 &  -37$^{\circ}$.6 & +7$^{\circ}$.2 & 2.2 $\times$ 10$^{-9}$ \\
\hline
\multicolumn{4}{l}{[a] Lapshov et al. 1992}\\
\end{tabular}
\end{center}
\end{table*}

\begin{figure}[!th]
    \resizebox{\hsize}{!}{\includegraphics{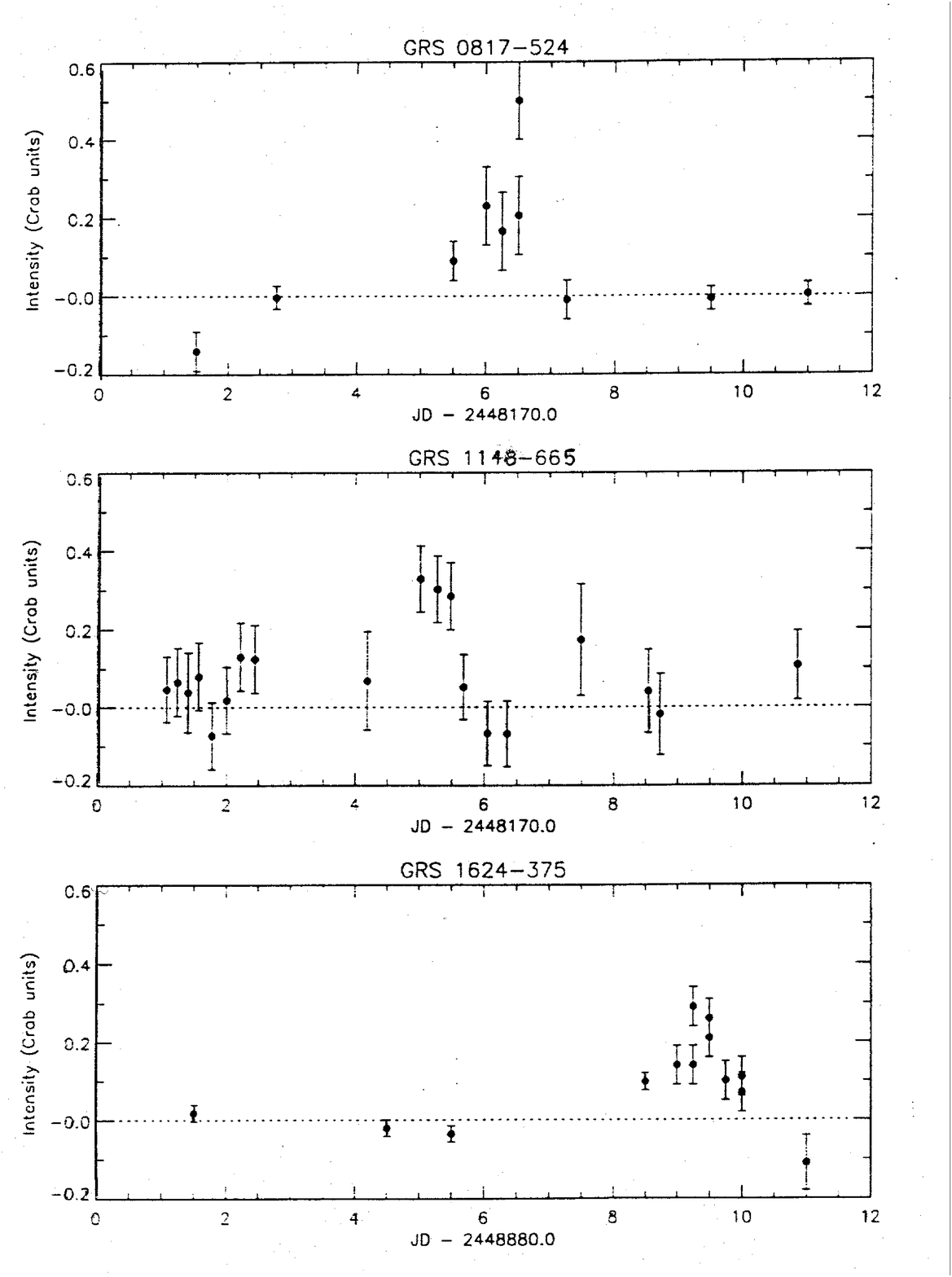}}
    \caption{The X-ray light-curves in the 8-15 keV of the three X-ray
             transients (lasting $\sim$ 1 day) discovered by WATCH on
             Granat and discussed in this paper.
             }
\end{figure}

\subsection{Search for correlated X-ray flares.}
We have searched for X-ray flares in the WATCH data base that
could have occurred in simultaneity to observed flares at other wavelengths 
reported elsewhere. The results were negative both for $\lambda$ Eridani 
(a flare was observed by {\it ROSAT} on 21 Feb 1991, Smith et al. 1993) and 
AU Mic 
(a flare was observed by {\it EUVE} on 15 July 1992, Cully et al. 1993). 
In the case of EV Lac, no WATCH data were available for an extraordinary 
optical flare that occurred on 15 Sep 1991 (Gershberg et al. 1992).

\section{Discussion}

\subsection{The fast X-ray Transients}
The three events quoted on Table 1 imply a rate of $\sim$ 5 year$^{-1}$. If 
the sources of the three events are galactic, their high lattitude would 
suggest that they are close to us, although no nearby known flare stars 
(Pettersen 1991) have been found in the corresponding WATCH error boxes. 

{\it GRS 1100-771.} 
Only four variable stars are catalogued in the WATCH error box: CS Cha, 
CT Cha, CR Cha and TW Cha. All are Orion-type variables. TW Cha undergoes 
rapid light changes with typical amplitudes of 1$^m$.
The whole error box lies in the Chamaeleon I star-forming cloud, and the 
position of 41 (out of $\sim$ 75) soft X-ray sources detected by {\it ROSAT}
in the cloud (Alcal\'a et al. 1995), are compatible with the position  
for GRS 1100-771 derived from the WATCH data. 
The DENIS near-IR survey of the region has revealed 170 objects in the cloud, 
mostly T Tau stars (Cambr\'esy et al. 1998).
Feigelson et al. (1993) also found that $\sim$80 \% of the X-ray sources are 
identified with T Tau stars with X-ray luminosities ranging from
6 $\times$ 10$^{28}$ to 2 $\times$ 10$^{31}$  \rm \ erg \ s$^{-1}$.

The WATCH detection implies a {\it flaring} luminosity 
L$_{x}$ = 0.6 $\times$ 10$^{30}$  (d/1 \rm \ pc)$^{2}$ \rm \ erg \ s$^{-1}$  
(8-15 keV),
or L$_{x}$ = 1.2 $\times$ 10$^{30}$ \rm (d/ \rm \ 1 pc)$^{2}$ \rm \ erg 
\ s$^{-1}$  (0.1-2.4 keV) for a thermal spectrum with kT = 10 keV and
no absorption (Greiner 1999).
If the X--ray source lies in the Chamaeleon I cloud (at $\sim$ 150 pc),
 L$_{x}$ = 1.35 $\times$ 10$^{34}$  \rm \ erg \ s$^{-1}$  (8-15 keV), or
 L$_{x}$ = 2.6 $\times$ 10$^{34}$  \rm \ erg \ s$^{-1}$  (0.1-2.4 keV).
with an integrated flare energy of $\sim$ 10$^{38}$ erg across the 
X-ray band (0.1-10 keV). 
Taking into account that kT $\sim$ 10 keV, a value higher than that observed 
in the T Tau star V773 (Tsuboi et al. 1998), and that the energy release would
be $\sim$ 100 times larger than that seen in the YSO IRS 15 in the $\rho$ Oph 
cloud (Grosso et al. 1997), we tentatively suggest a superflare arising from 
one of the YSOs in the Chamaeleon I star-forming 
cloud as the most likely candidate for GRS 1100-771. 
If this is indeed the case, it will be difficult to explain how this 
exceptionally high X--ray luminosity can be explained by a solar-like
coronal emission mechanism. However, a large amount of e\-nergy, stored in 
large magnetic structures that it is set free by magnetic reconnection events 
(see Grosso et al. 1997), can a\-ccount for the WATCH detection.
But we notice that the highest quiescent luminosity for any of the X-ray 
sources detected by {\it ROSAT} in the WATCH error box is 
L$_{x}$ $\sim$ 10$^{31}$  \rm \ erg \ s$^{-1}$  (0.1-2.4 keV) at the utmost 
(Alcal\'a et al. 1997), i.e. 100 times lower than that seen in IRS 43 for
which a steady supply of energy (like a large number of reconnection events)
was proposed (Preibisch 1998).
In any case, the possibility of GRS 1100-771 being an isolated neutron star 
unrelated to the cloud cannot be excluded, given the relatively large error 
box provided by WATCH.

{\it GRS 2037-404.} It was discovered on 23 Sep 1992 (Castro-Tirado, Brandt 
\& Lund 1992), and lasted for 110 min, reaching a peak intensity of $\sim$ 
0.8 Crab. 
On our request, a Schmidt plate was taken at La Silla on 29 Sep, and based on 
this plate, it was reported the presence of the Mira star U Mic near maximum 
brightness at 7.0 mag (Della Valle \& Pasquini, 1992). U Mic is a Mira star 
and we do not consider this star as a candidate due to the different timescales
of the involved physical processes. Moreover it is too far away from the error 
box centre ($\sim$ 2$^{\circ}$). Hudec (1992) proposed another variable 
object inside the WATCH error box as a candidate to be further investigated. 
This is the star RV Mic (B $\sim$ 10.5) discovered in 1948 (Hoffmeister 1963). 
Although is classified as a Mira Cet type star, it is apparently a highly 
variable object. Another four variables are within the 1$^{\circ}$ radius 
WATCH error box: UU Mic (a RR Lyr type), UW Mic (a semi-regular pulsating 
star), 
SZ Mic (an eclipsing binary) and UZ Mic (an eclipsing binary of W UMa type). 
In any case, we carefully examined the Schmidt plate and found no 
object inside the 1$^{\circ}$ radius  WATCH error box that would have varied 
by more then 0.5$^m$ when comparing with the corresponding ESO Sky 
Survey plate.

{\it GRS 2220-150.} No variable stars are catalogued within the WATCH error 
box. The high flux (2.1 Crab) is similar to the peak of the 
fast transient at 20h14m + 30.9 discovered by {\it OSO-8} in 1977 
(Selermitsos, 
Burner \& Swank 1979). The peak luminosity for GRS 2220-150 was L$_{x}$  
 = 1.5 $\times$ 10$^{30}$ \rm \ (d/1\ pc)$^{2}$ \ erg\ s$^{-1}$ (8-15 keV), 
or  L$_{x}$ =  6 $\times$ 10$^{30}$ \rm \ (d/1\ pc)$^{2}$ \rm \ erg \ 
s$^{-1}$  (2-15 keV) if we assume a Crab-like spectrum. 

Considering the high peak luminosities, a flare of a dMe or a dKe star may be 
excluded in these two latter cases. No nearby RS CVn stars from the list of 
Lang (1992) were found in the error boxes. It can be possible that these 
events would be associated to old isolated neutron stars accreting 
interstellar matter with unstable nuclear burning (Ergma \& Tutukov 1980). 
According to Zduenek et al. (1992), when the accretion rate is higher than 
10$^{13}$ \rm \ g \ s$^{-1}$, 
the hydrogen burning triggered by electron capture becomes unstable. As the 
mass of the accreted envelope should be 10$^{23}$ g, several hundred 
years will be required for accreting this ammount of matter. Assuming that the 
number of isolated neutron stars with such high rate is 10$^{5}$ (Blaes \& 
Madau 1993), one should expect 10-100 fast X-ray transients per year. 

\subsection{The longer duration X-ray Transients}

With the exception of GRS 1133+54, the other three sources quoted in 
Table 2 are concentrated near the galactic plane,
suggesting that they could be more distant than the three fast X-ray 
transients mentioned above. In this case, no nearby dwarf star or RS CVn 
systems have been identified within the corresponding error boxes. 

{\it GRS 0817-524.} No variable stars are catalogued in the error box. 
However, we note the presence of an X-ray source detected by {\it ROSAT}:
1RXS J081938.3-520421 is listed in the All-Sky Bright Source
Catalogue (Voges et al. 1996), with coordinates  R.A.(2000) = 8h19m38.3s, 
Dec(2000) = -52$^{\circ}$ 04' 21". 

{\it GRS 1148-665.} Two variable stars lie within the error box: 
CY Mus (a RR Lyr type) and TW Mus (an eclipsing binary of W UMa type).
We also note here the existence of a {\it ROSAT} All-Sky bright source:
1RXS J115222.9-673815, at R.A.(2000) = 11h52m22.9s, 
Dec(2000) = -67$^{\circ}$ 38' 15".   

{\it GRS 1624-375.} No variable stars are reported within the error
box.  A 300-s optical spectrum taken in March 1997 at the 2.2-m ESO 
telescope of the possible candidate suggested by Castro-Tirado (1994) reveals 
that this is a M-star unrelated to GRS 1624-375 (Lehmann 1998). 
However there are three sources in the {\it ROSAT} Catalogue: 
1RXS J162730.0-374929, 1RXS J162751.2-371923 and
1RXS J163137.0-380439.

Taking into account the {\it a priori} 
probability of finding such bright X-ray sources in the typical WATCH 
error boxes ($\sim$ 1.5 sources per error box),  
we conclude that none of the 1RXS sources is
related to the longer duration X-ray Transients detected by WATCH.
   
In the {\it Ariel V} Catalogue of fast X-ray Transients, Pye \& McHardy (1983)
described 10 X-ray transients with durations ranging  0.5-4 days. Some of them 
were associated with known Be-neutron stars systems, like 4U 0114+65, and 
other with RS CnV systems, like $\sigma$ Cen or DM UMa (for which a flare 
lasting 1.5 days was observed by HEAO-2). The {\it RossiXTE} satellite has 
recently detected other two X-ray transients: XTE J0421+560 and XTE J2123-058.
XTE J0421+560 lasted for $\sim$ 2 days (Smith et al. 1998) and it is probably 
related to a black hole in a binary system. XTE J2123-058 lasted for $\sim$ 
5 days and is presumably related to a low-mass binary in which a neutron star
undergoes type-I bursts (Levine et al. 1998, Takeshima and Strohmayer 1998). 
Another short-duration event, lasting $\sim$ 2 days has
been observed in X-rays and radio wavelengths in the superluminal galactic 
transient GRS 1915+105 (Waltman et al. 1998).
We consider that the three long duration events reported 
in this paper, with fluxes in the range 0.35--0.5 Crab assuming a Crab-like
spectrum, are likely to have originated from compact objects in  
low-mass or high-mass X-ray binaries, similarly to XTE J0421+560
and XTE J2123-058.

\section{Conclusions}

Amongst the 3 fast X-ray transients discovered by WATCH in 
1990-92, GRS 1100-771 might be related 
to a superflare arising from a YSO in the Chamaeleon I cloud, that would 
make -to our knowledge- the strongest flare ever seen in such an object. 
But the possibility of GRS 1100-771 being an isolated neutron star unrelated 
to the cloud cannot be excluded, given the relatively large error box provided
by WATCH.
Amongst the other 3 longer duration X-ray transients, none of them seem to 
be related to known objects, and we suggest that the latter are likely to 
have originated from compact objects in low-mass or high-mass X-ray binaries.


\vspace{1cm}

\begin{acknowledgements}
We would like to thank the referee, Thi\'erry Montmerle, for his very 
constructive comments. We are grateful to J. Greiner for useful discussions 
and for providing the expected fluxes in the 0.1-2.4 keV by means of the EXSAS 
package. We also thanks R. Hudec for valuable comments, and G. Pizarro, 
A. Smette and R. West for the Schmidt plates taken at ESO \' \rm s La Silla 
Observatory, and to I. Lehmann for the optical spectrum taken at the 2.2-m 
ESO telescope in the GRS 1624-375 field.
We are indebted to the staff of the Evpatoria ground station in Ukraine 
and those of the Lavotchkin and Babakin Space Center in Moscow, where 
{\it Granat} was built. 
This search made use of the SIMBAD database, at
the Centre de Donn\'ees astronomiques de Strasbourg. This work has been 
partially supported by Spanish CICYT grant ESP95-0389-C02-02.
\end{acknowledgements}

\end{document}